\documentclass[conference]{IEEEtran}
\IEEEoverridecommandlockouts
\usepackage{cite}
\usepackage{amsmath,amssymb,amsfonts}
\usepackage{algorithmicx}
\usepackage{graphicx}
\usepackage{textcomp}
\usepackage{xcolor}
\usepackage{hyperref}
\usepackage{algorithm2e}
\usepackage{algpseudocode}
\usepackage{tikz}
\usepackage{comment}
\usepackage{cleveref}
\usepackage{booktabs}
\usepackage{boxedminipage}
\usepackage{atbegshi}

\usetikzlibrary{arrows.meta, positioning, fit}

\def\BibTeX{{\rm B\kern-.05em{\sc i\kern-.025em b}\kern-.08em
    T\kern-.1667em\lower.7ex\hbox{E}\kern-.125emX}}
\begin{document}

\title{Decentralized Proof-of-Location for Content Provenance: Towards Capture-Time Authenticity%
\thanks{\fontsize{7}{8}\selectfont This work was supported in part by the 2025/2026 Charlemagne Prize Academy Fellowship of the Charlemagne Prize Foundation.}}

\author{
\IEEEauthorblockN{
Eduardo Brito\IEEEauthorrefmark{1}\IEEEauthorrefmark{2},
Fernando Castillo\IEEEauthorrefmark{3},
Amnir Hadachi\IEEEauthorrefmark{2},
Ulrich Norbisrath\IEEEauthorrefmark{2},
Jonathan Heiss\IEEEauthorrefmark{3}
}
\IEEEauthorblockA{\IEEEauthorrefmark{1}Cybernetica AS, Estonia \quad \IEEEauthorrefmark{2}University of Tartu, Estonia \quad \IEEEauthorrefmark{3}TU Berlin, Germany}
\IEEEauthorblockA{
eduardo.brito@cyber.ee, \{amnir.hadachi,ulrich.norbisrath\}@ut.ee, \{fc,jh\}@ise.tu-berlin.de
}
}

\AtBeginShipoutFirst{%
\begin{tikzpicture}[remember picture, overlay]
  \node[
    anchor=south,
    yshift=0.7cm,
    draw,
    inner sep=6pt,
    text width=0.7\paperwidth
  ] at (current page.south) {%
    \textbf{Preprint.} This work has been accepted for publication at the 5th International Workshop on Architecting and Engineering Digital Twins (AEDT 2026), to appear in the Companion Proceedings of the 23rd IEEE International Conference on Software Architecture (ICSA 2026).
  };
\end{tikzpicture}
}

\maketitle

\begin{abstract}
Reliable use of real-world data requires confidence that recorded evidence reflects what actually occurred at the moment of capture. In adversarial or incentive-misaligned cyber-physical settings, device-centric provenance and post-capture verification are insufficient to provide that guarantee. This paper builds on Proof-of-Location (PoL) as a baseline for establishing where and when events take place, and extends it with a witnessing-zone architecture in which multiple independent observers collectively validate physical events. The resulting approach produces auditable evidence artifacts that can support downstream systems in cyber-physical settings, without relying on centralized trust. Through representative scenarios and simulation-based evaluation, this paper shows how such architectures improve sensor data trustworthiness and resilience to fabricated or staged events.
\end{abstract}

\begin{IEEEkeywords}
Proof-of-Location; Content Provenance; Capture-Time Evidence; Cyber-Physical Systems
\end{IEEEkeywords}

\section{Introduction}
In distributed, multi-stakeholder cyber-physical settings with externally sourced observations, the question is not only whether data was recorded, but whether it can be trusted as a faithful account of what physically occurred. When incentives to misrepresent events exist, such trust cannot rest on a single device or party. This challenge is especially relevant for Digital Twins (DTs) that ingest externally sourced observations into state updates, analytics, and audit trails. Decentralized Proof-of-Location (PoL) systems provide a principled foundation. By combining distance-bounding protocols with Byzantine fault-tolerant consensus among independently positioned witnesses, PoL architectures produce cryptographic proofs that a device was present within a bounded spatial region during a bounded temporal interval~\cite{Brito2025SciReports,brito2025taxonomy}. These proofs establish \emph{time-and-space integrity}: verifiable evidence of where and when an observation was captured, grounded in physical-layer constraints.

However, time-and-space integrity is insufficient when the content of a record must also be trusted. A device with a valid location proof can photograph synthetic imagery, record staged environments, or transmit misleading telemetry while satisfying all spatial and temporal constraints~-- an attack class we term \emph{scene spoofing}. Single-device provenance frameworks such as C2PA~\cite{C2PA2024Spec} provide post-capture chain-of-custody guarantees, but inherit the device’s unverified claim about what it observed, leaving scene spoofing unaddressed.

This paper introduces an augmented witnessing-zone architecture that extends decentralized PoL with \emph{capture-time evidence}: attestations binding digital records not only to place and time, but also to physical context independently observed by a quorum of witnesses at capture time. The contribution is an architectural extension that adds multimodal witness sensing, policy-driven local reasoning, and a compact evidence object that downstream systems can verify before accepting real-world data as input. In this model, a claim is admitted only when a quorum validates both spatio-temporal proximity and contextual coherence within a single interval. This trust layer is particularly relevant for DTs when external observations influence twin-state evolution, but the architecture is intended for broader use across distributed, multi-stakeholder cyber-physical settings. We motivate the design through three use-case domains~-- agri-commodity containerization, media provenance, and smart mobility~-- and evaluate it through domain-specific instantiation and discrete-event simulation quantifying admission rates, rejection of distance-fraud and scene-spoofing attempts, and boundary behavior under stochastic channel conditions.

The remainder of the paper is as follows. Section~\ref{sec:use-cases} presents the use cases. Section~\ref{sec:related-work} surveys related work. Section~\ref{sec:witnessing-zone-baseline} specifies the baseline model. Section~\ref{sec:architecture} introduces the augmented architecture. Section~\ref{sec:evaluation} presents the evaluation. Section~\ref{sec:discussion} discusses the paper and Section~\ref{sec:conclusion} concludes.

\section{Use Cases}
\label{sec:use-cases}
The following three domains share a common structure: multiple parties with misaligned incentives must later agree on what occurred at a physical event, based on digital records created at capture time. Physical environments are typically dense with ambient signals that collectively encode contextual state. Yet conventional provenance mechanisms bind records to a device, not to this surrounding sensory field. In each domain, time-and-space integrity alone leaves a gap that scene spoofing can exploit.

\paragraph{Agri-Commodity Containerization}
Due-diligence regimes such as the EU Deforestation Regulation require traceable linkages between farm-level geolocation and downstream events including lot sealing, warehouse ingress or egress, and container loading~\cite{EU_Regulation_2023_1115}. Along this chain~-- of exporters, aggregators, shipping agents, importers, and auditors~-- economic incentives diverge. Operational bottlenecks such as re-stows or subcontracted loading further complicate reliable recording of what was handled and when~\cite{li2025critical,zhen2013review}.
A device proven to be at a loading bay during a sealing window does not establish which lot was inside the container, whether a seal photograph depicts the actual seal, or whether pallet sequences match the physical shipment. These scene-spoofing vectors satisfy proximity constraints while fabricating evidentiary content. Nearby infrastructure could sense BLE signatures of tagged pallets, ambient temperature, or short-range visual confirmation of seal presence, yet cannot verify sealed contents. Structural limits of third-party audits leave such discrepancies undetected~\cite{lebaron2016ethical,searcy2024auditor}. Similar patterns arise under the U.S.\ DSCSA~\cite{USFDA_DSCSA_Act}, EPA e-Manifest~\cite{USEPA_2025_eManifest}, and fisheries traceability regimes~\cite{FAO_2024_SeafoodTraceability}. Because disputes may surface months later, records must remain self-contained and independently verifiable.

\paragraph{Media Provenance}
In conflict covering, electoral, or breaking-news settings, creators, platforms, newsrooms, and subjects hold distinct stakes in whether footage is accepted as authentic. Synthetic media reduces fabrication costs, while platform processing often strips metadata anchoring assets to specific times and locations~\cite{ChesneyCitron2019DeepfakesLaw,Tolosana2020DeepfakesSurvey,IPTC_2013_MetadataDeletion}. The resulting ``liar's dividend'' benefits both fabricators and those dismissing genuine evidence~\cite{schiff2022liar}.
A camera with a valid location proof can still record a staged scene or a screen displaying synthetic content. Surrounding infrastructure could capture ambient audio fingerprints, visual scene descriptors, or RF environments to corroborate the presence of a real scene, but cannot attest to narrative truthfulness. Reactive forensics detection methods degrade under transcoding and adversarial adaptation~\cite{DARPA_2024_SemaFor,NIST2024AI1004}. Whether verification occurs within hours or years, stakeholders require evidence tied to physical scene conditions at capture, not merely to device provenance.

\paragraph{Smart Mobility}
Urban mobility governance increasingly relies on where-when rules such as cordon charges, low-emission zones, and geofenced pickup areas. When contested, consequences diverge across drivers, fleet operators, municipalities, and insurers. The evidentiary substrate is fragmented: fixed cameras have blind spots, GNSS drifts in urban canyons, and telemetry pipelines introduce synchronization inconsistencies~\cite{weng2023characterization,liu2021smartphone}.
For pure presence queries~-- e.g., whether a vehicle entered a charge polygon during a billing interval~-- time-and-space integrity may suffice. Many disputes, however, concern what occurred there: whether a vehicle was stationary or transiting, whether a pickup occurred at the designated curb, or whether a photographed plate matches the vehicle present. Incentives to misrepresent are direct. Roadside infrastructure may sense RF fingerprints, Doppler profiles, or image hashes, yet cannot establish driver identity or intent. Privacy constraints further limit persistent camera-based repositories~\cite{bogdanov2025zero,TfL_2025_ULEZReporting}. As appeals may arise weeks later, evidence must remain portable and verifiable across institutional boundaries.

\section{Related Work}
\label{sec:related-work}
\paragraph{Sensor- and Content-Level Authentication and Provenance}
A wide range of techniques support authentication and provenance across sensing modalities. Device-side signing embeds cryptographic credentials at capture, enabling later verification of integrity and origin, as formalized by the Content Authenticity Initiative and C2PA standards~\cite{CAI2025Credentials,C2PA2024Spec,C2PA2024Attestation}. Watermarking provides complementary, content-embedded signals but faces trade-offs between robustness and removability, as demonstrated by both resilient schemes and removal attacks~\cite{Wen2023TreeRing,Zhao2024WatermarkRemoval,NIST2024AI1004}. Reactive forensic methods, such as device fingerprints (e.g., camera PRNU), remain useful yet fragile against sophisticated adversaries~\cite{Lukas2006PRNU,MartinRodriguez2023PRNUStress}. Collectively, these approaches strengthen integrity guarantees but assume the capture device is trustworthy and that scene conditions were not spoofed.

\paragraph{Threat Landscape: Generative Manipulation and Scene Spoofing}
Advances in generative AI models have reduced the cost of fabricating or altering sensor-derived content, undermining confidence in digital evidence and amplifying the “liar's dividend,” i.e., the strategic dismissal of genuine recordings as fake~\cite{Tolosana2020DeepfakesSurvey,Mirsky2021DeepfakesSurvey,Verdoliva2020MediaForensics,ChesneyCitron2019DeepfakesLaw}. As detectors improve, adversaries adapt, rendering reactive forensics a moving target. This dynamic motivates proactive authenticity: binding acquisitions to verifiable spatial and temporal context at capture, so provenance becomes cryptographic evidence rather than post-hoc probabilistic inference~\cite{NIST2024AI1004}. In particular, scene spoofing~-- e.g., recording staged environments or photographed screens~-- emerges as a first-class threat that single-device assurances cannot exclude.

\paragraph{Cryptographic and Infrastructural Approaches to Trustworthy Sensing}
Proactive authenticity increasingly combines cryptography with distributed infrastructure. Anchoring content hashes and manifests on distributed ledgers yields tamper-evident, auditable records across capture and publication, complementing CAI/C2PA and Project Origin~\cite{Ramachandran2018BlockchainProvenance,ProjectOrigin2025,C2PA2024Spec}. Hardware-backed attestations, such as secure enclaves signing sensor outputs within capture pipelines, strengthen device integrity but still concentrate trust in a single endpoint~\cite{Naveh2016PhotoProof,liu2022vronicle,mesa2025enabling,C2PA2024Attestation}. To address scene spoofing and coordinated relays, emerging work shifts toward collective attestation, where multiple independent devices corroborate the same event, distributing trust across a witness set~\cite{Brito2025SciReports}.

\paragraph{Proof-of-Location (PoL) Protocols and Architectures}
Proof-of-Location (PoL) systems cryptographically establish where and when presence occurred. Typical designs combine short-range distance bounding with signed witness attestations, aggregated or ledger-anchored for immutability and auditability~\cite{brito2025taxonomy}. By fusing physical signal constraints with cryptographic commitments, PoL mitigates spoofing, relay, and replay attacks that affect conventional positioning systems. Decentralized PoL distributes trust across quorum-based witness sets, preventing unilateral proof fabrication~\cite{FOAM2018Whitepaper,Brito2025SciReports}. This provides a natural substrate for grounding digital records in real-world spatio-temporal conditions.

\paragraph{Bridging Spatial Proofs and Multimodal Provenance}
Extending PoL to sensor provenance enables capture-time certificates coupling spatial, temporal, and environmental evidence. Nearby devices may co-sign attestations incorporating ambient features~-- light, sound, radio context~-- forming multimodal proofs that bind content to its physical surroundings~\cite{Brito2025SciReports,Castillo2025DAMs}. As sensor networks evolve toward reasoning-capable agents, they can validate contextual coherence across modalities, strengthening collective attestations~\cite{Castillo2025DAMs}. The objective is to make synthetic fabrications increasingly costly to reproduce coherently across independent observers, while genuine captures remain anchored to shared spatio-temporal state and auditable records~\cite{Ramachandran2018BlockchainProvenance,C2PA2024Spec}. This convergence defines the foundation of the present work.

\section{The Witnessing Zone: Baseline Model}
\label{sec:witnessing-zone-baseline}
\begin{figure}[htbp]
    \centering
    \includegraphics[width=\columnwidth]{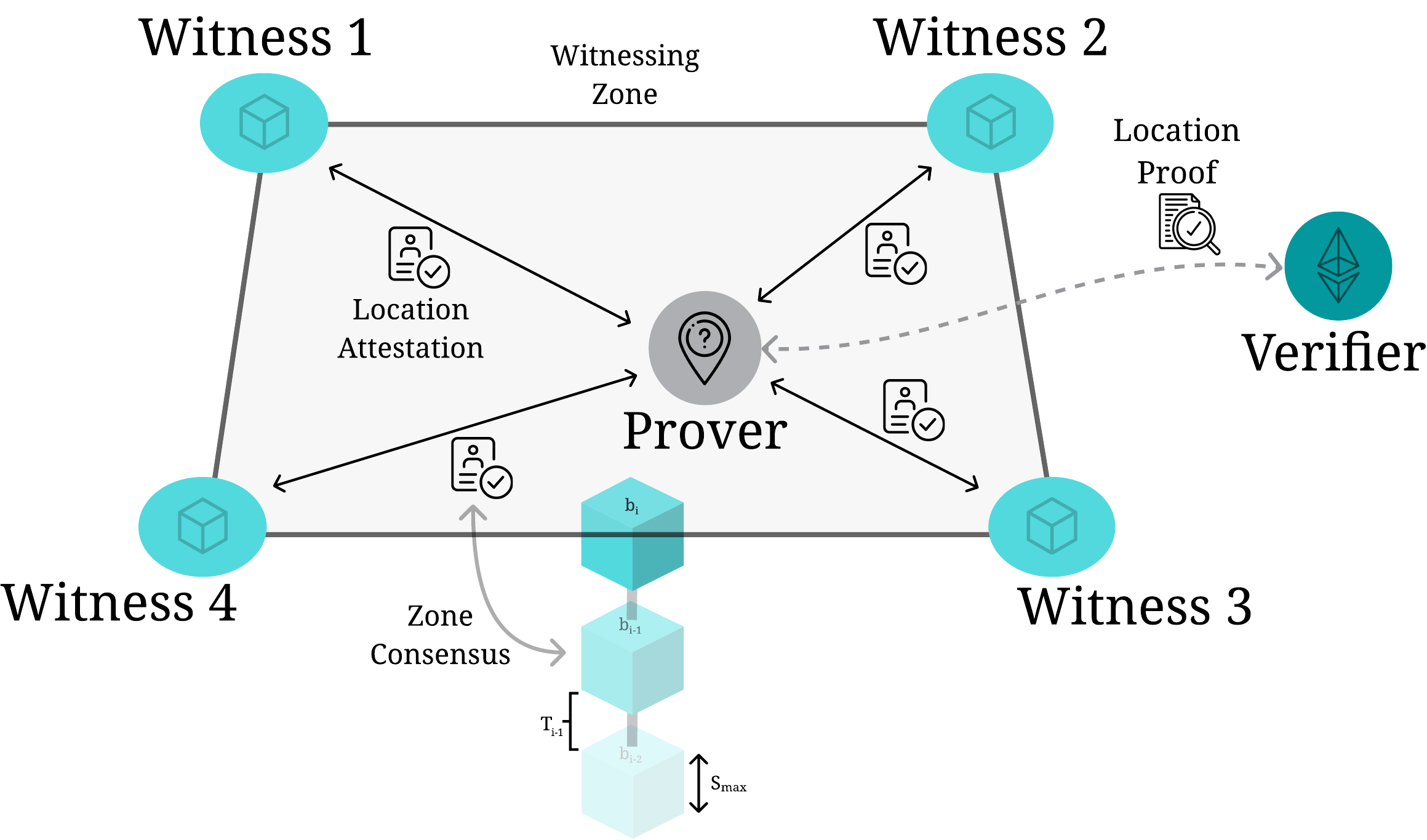}
    \caption{Baseline witnessing zone model. Witnesses form a synchronized mesh, attest to a prover's location, and collectively produce a location proof verified by an external verifier.}
    \label{fig:witnessing-zone-model}
\end{figure}

Following advancements on decentralized PoL systems, we adopt the \emph{witnessing zone} as the baseline architectural model for collective location attestation under timing and trust constraints~\cite{brito2025taxonomy,Brito2025SciReports}. This section specifies the minimal structure and guarantees of a witnessing zone (Figure~\ref{fig:witnessing-zone-model}), excluding capture-time contextual sensing and local reasoning, which are introduced later.

\paragraph{Zone Structure} A witnessing zone is a bounded, single-hop spatial region populated by a finite set of authenticated witnesses operating under a shared protocol. Witnesses are fixed-position nodes forming a fully connected, non-hierarchical mesh. Zone operation is discretized into fixed-length block intervals of duration \(T\). All witness attestations are bound to a specific interval, and only attestations produced within the same interval are eligible for quorum formation, enforcing temporal soundness and replay resistance~\cite{brito2025taxonomy}. For each interval, witnesses execute a Byzantine fault-tolerant consensus protocol to finalize the set of claims admitted by quorum, defining a consensus round whose outcome is a finalized and hashed block \(b_i\), containing all claims admitted by quorum during that interval, yielding a totally ordered, append-only, tamper-evident event log~\cite{Brito2025SciReports}. Each witness maintains a local replica of this ordered log and associated zone configuration state. An operational information management layer governs zone membership, key rotation, configuration updates, and retention policies, and may optionally anchor block hashes or zone state commitments to an external public ledger for additional auditability. In standard two-dimensional deployments, zones are instantiated with four coplanar witnesses and a quorum threshold of \(k=3\), tolerating one Byzantine witness.

\paragraph{Interaction Model} Within each interval, a prover broadcasts a location claim. Each witness independently performs a time-critical physical-layer protocol with the prover, typically distance bounding, to derive an upper bound on physical proximity. Based on protocol validity and proximity alone, witnesses decide whether to attest. A claim is admitted if and only if at least \(k\) distinct witnesses produce valid attestations within the same interval.

\paragraph{Identity Model} Each witness possesses a stable cryptographic identity represented by a public–private key pair. Witness identities are established prior to operation and may be realized via PKI-based certification, consortium-managed registries, or ledger-anchored identity records, provided identities are authenticated and non-forgeable. Provers are treated as untrusted and may be ephemeral; prover identity is not relied upon for correctness. All attestations are scoped to a unique zone id \(Z_z\), binding them to a specific witness set and configuration.

\paragraph{Security Guarantees} The baseline witnessing zone provides spatio-temporal soundness, meaning that the protocol admits only claims that satisfy time-and-space integrity: presence within a bounded spatial region during a bounded temporal interval. Distance bounding mitigates relay and replay attacks, while quorum and Byzantine consensus ensure fault tolerance and global consistency under the assumption that fewer than \(k\) witnesses are compromised. The model does not assess semantic correctness of claims or environmental coherence; it proves that a claim was collectively accepted at a given place and time, but not \emph{why} beyond proximity and quorum.

\section{Augmented Witnessing Zone Model}
\label{sec:architecture}
\begin{figure*}[t]
    \centering
    \includegraphics[width=\textwidth]{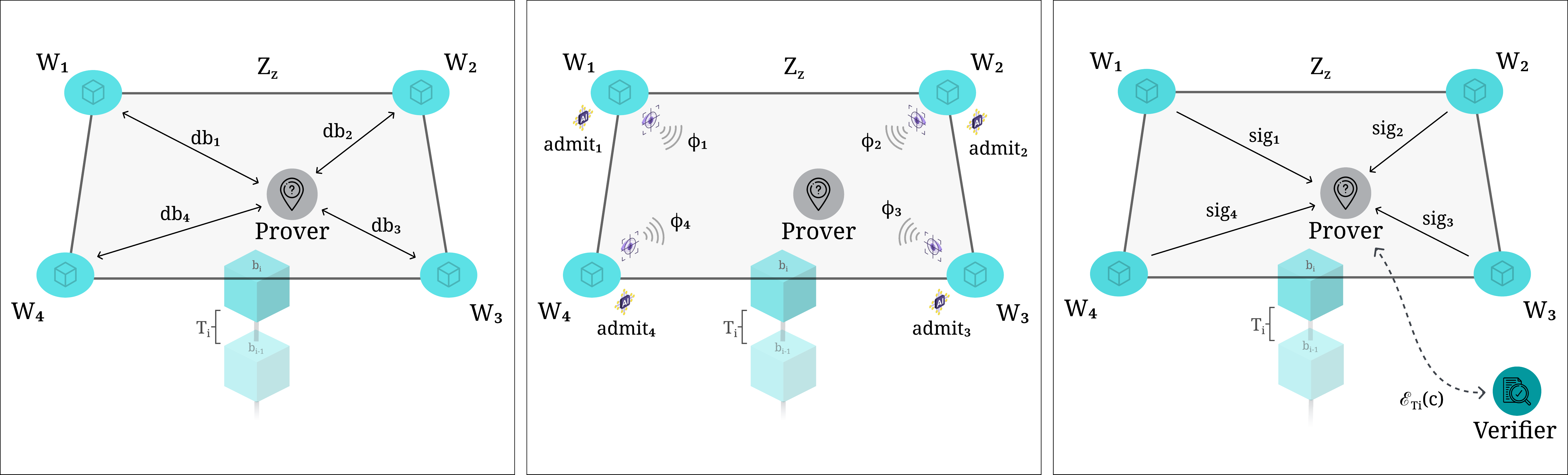}
    \caption{Augmented witnessing zone architecture. After distance bounding, each witness executes local sensing and policy-based reasoning before contributing to quorum formation.}
    \label{fig:augmented-witnessing-zone}
\end{figure*}

Building on the baseline witnessing zone model, we augment the witness role beyond proximity verification by incorporating capture-time contextual sensing and local reasoning into the per-interval witness decision process. The contribution is an architectural augmentation: it preserves the geometric configuration, block-interval timing, identity assumptions, quorum thresholds, and Byzantine consensus foundations of the base PoL model while refining what each witness evaluates before contributing to quorum. These additions operate at the level of the \emph{witness machine}, i.e., the local decision logic executed independently by each witness during a block interval. As a result, claims remain admitted through interval-bound quorum and consensus, but witness attestations now jointly reflect spatio-temporal proximity and contextual coherence evaluated at capture-time (Figure~\ref{fig:augmented-witnessing-zone}).

\subsection{Capture-Time Sensing and Local Reasoning}

Each block interval \(T_i\) begins with the prover broadcasting a context-bearing claim \(c \in \mathcal{C}\), consisting of a structured representation of the asserted event and an optional set of disclosed features \(\phi_P\). Within the same interval, each witness executes the steps summarized in Algorithm~\ref{alg:witness-machine}.

\paragraph{Distance Bounding and Sensing}
Each witness \(w_j\) engages in a time-critical exchange with the prover to derive an upper bound \(db_j\) on physical distance. In parallel, the witness samples its configured sensing modalities (e.g., RF, visual, audio, inertial, ambient) and extracts feature descriptors \(\phi_j = \{f_j^{(m)}\}_m\), annotated with quality and freshness metadata. Feature descriptors are compact, policy-relevant representations of raw sensor signals. For example, under a BLE-based policy, \(f_j^{(\text{BLE})}\) may be a beacon fingerprint vector or similarity score, whereas under a visual policy, \(f_j^{(\text{vis})}\) may be an embedding vector or perceptual hash used to compute a thresholded similarity. 

\paragraph{Policy-Based Reasoning and Evidence Commitment}
Each witness evaluates a versioned policy predicate
\[
\mathsf{Admit}_v(c, \phi_P, \phi_j, db_j) \rightarrow \{0,1\},
\]
which may encode deterministic rules, statistical thresholds, or learned comparison models. The internal computation need not be externally reproducible; however, all inputs influencing the decision are cryptographically committed. If the predicate evaluates to \(\mathsf{admit}_j = 1\), the witness commits its local execution state to a Merkle root \(R_j\) and signs the tuple \((b_i, c, R_j, v, Z_z)\).

Quorum formation is strictly interval-bound: only signatures produced within the same block interval \(T_i\), anchored in block \(b_i\), are composable. If at least \(k\) witnesses produce valid signatures, the claim is admitted by the zone and passed to evidence assembly.

\begin{algorithm}[t]
\fbox{%
\begin{minipage}{\dimexpr\columnwidth-1.2em\relax}
\KwIn{Claim $c$, disclosed features $\phi_P$, policy version $v$}

Derive distance bound $db_j$ via distance-bounding protocol\;
Sample sensors and extract features $\phi_j$\;
$\mathsf{admit}_j \leftarrow \mathsf{Admit}_v(c, \phi_P, \phi_j, db_j)$\;

\If{$\mathsf{admit}_j = 1$}{
  Commit inputs and outputs to Merkle root $R_j$\;
  $sig_j \leftarrow \mathsf{Sign}_j(b_i, c, R_j, v, Z_z)$\;
  Broadcast signature to zone\;
}
\end{minipage}%
}
\vspace{0.6em}
\caption{Witness Machine Execution for Claim $c$ in Block Interval $T_i$}
\label{alg:witness-machine}
\end{algorithm}

\subsection{Evidence Semantics and Verification Interface}

Once a quorum of at least \(k\) witnesses has produced valid signatures for the same claim \(c\) within interval \(T_i\), the witnessing zone assembles a verifiable \emph{evidence object}. This object is the sole externally visible artifact of the augmented protocol and serves as the unit of verification for downstream systems.

The evidence object does not re-execute or reinterpret witness reasoning. Instead, it binds the claim to the set of committed witness decisions and the policy under which they were made:
\[
\mathcal{E}_{T_i}(c) :=
\big\langle
b_i,\;
c,\;
\{R_j\}_{j \in Q},\;
\sigma_Q,\;
v,\;
Z_z
\big\rangle,
\]
where \(Q\) denotes the quorum of admitting witnesses (\(|Q| \geq k\)) and \(\sigma_Q\) is the aggregated (threshold) signature.

\begin{figure}[htbp]
    \centering
    \includegraphics[width=\columnwidth]{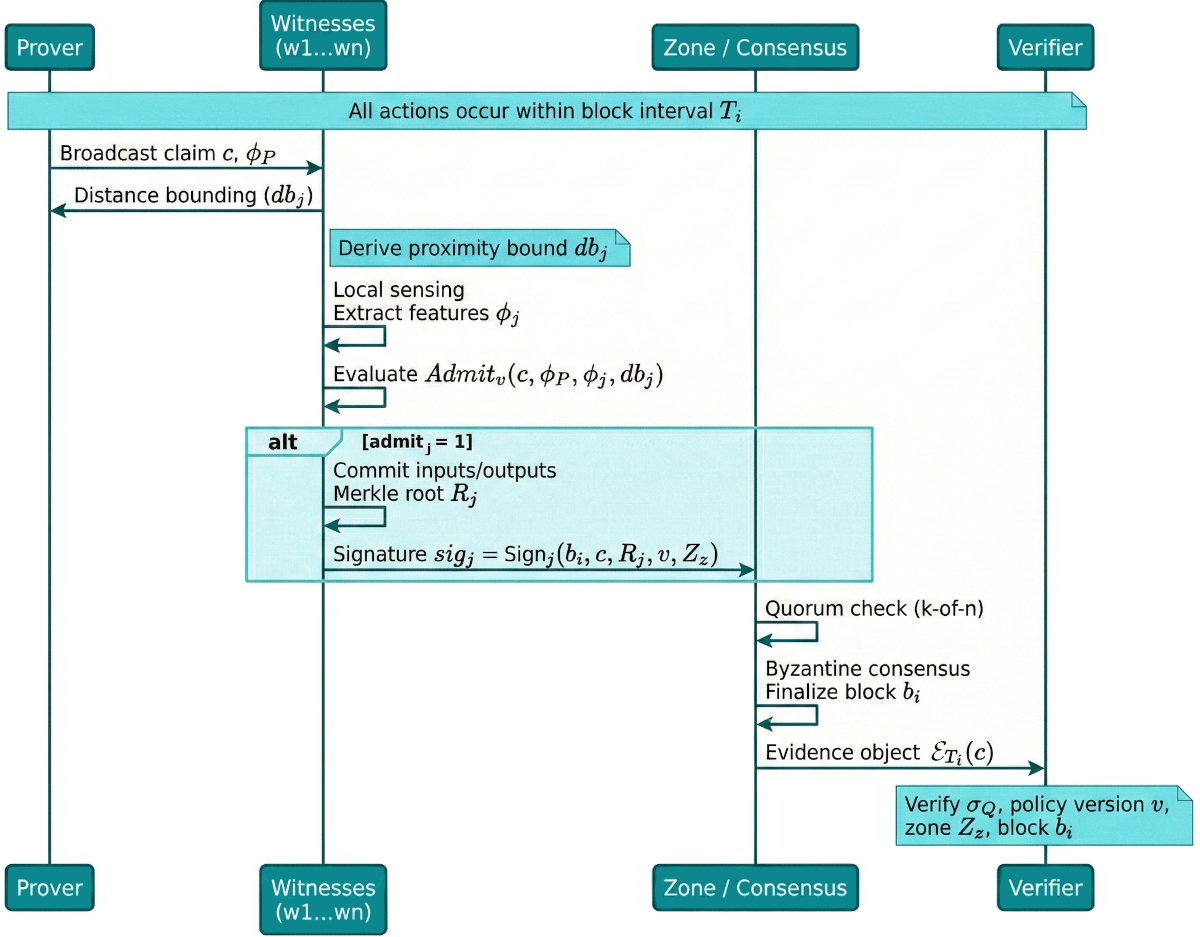}
    \caption{Interval-bound interaction flow of the augmented witnessing zone.}
    \label{fig:augmented-sequence}
\end{figure}

The semantics of \(\mathcal{E}_{T_i}(c)\) assert that (i) the claim was evaluated within a bounded execution window, (ii) a quorum of independent witnesses satisfied both distance-bounding constraints and the policy predicate, and (iii) each witness irrevocably committed to the inputs and outputs of its local reasoning. Verification checks \(\sigma_Q\) against the zone public keys and validates the policy version \(v\). Depending on the trust model, verifiers may accept the evidence object directly or request selective opening of committed elements. This makes \(\mathcal{E}_{T_i}(c)\) a compact interface for downstream consumers, including provenance services and DT data pipelines, to decide whether a real-world observation is admissible for state updates, analytics, or later dispute resolution. Figure~\ref{fig:augmented-sequence} summarizes the interval-bound interaction flow of the augmented witnessing zone.

\subsection{Deployment Profiles and Policy Examples}

The augmented protocol generalizes across deployments with different operational constraints, sensing capabilities, and evidentiary requirements. In all cases, distance bounding remains the core spatio-temporal guardrail, while additional sensing and reasoning are applied only after proximity has been validated. We highlight three representative profiles, Cocoa, Media, and Mobility, that illustrate how the architecture can be instantiated in practice.

\paragraph{Cocoa (Supply Chain Sealing and Hand-off)}
In this scenario, the prover is a container-mounted device at the point of sealing or custody transfer (e.g., at a port or cooperative facility), and the witnesses are fixed infrastructure nodes positioned at known locations near the hand-off point. The primary objective is to bind a transfer event to its physical location and time, ensuring that hand-offs occur \emph{in situ} and are witnessed by multiple independent parties.
\begin{itemize}
    \item \textbf{Sensors:} BLE signal strength, ambient temperature, short-range ultrasonic chirps.
    \item \textbf{Policy (v1):} Admit if \(k\)-of-\(n\) (i) distance bounds are valid, and (ii) the BLE fingerprint matches the known warehouse signature with similarity score above threshold \(\tau_{\text{BLE}} = 0.85\).
\end{itemize}

\paragraph{Media (Content Provenance in Public Environments)}
Here, the prover is a user device (e.g., smartphone or bodycam) broadcasting a claim over media being recorded or streamed, and the witnesses are surrounding IoT nodes (e.g., streetlamps, kiosks, or vehicle-mounted sensors). The aim is to tie digital media to its spatial and environmental context at capture time, in order to produce verifiable provenance certificates.
\begin{itemize}
    \item \textbf{Sensors:} visual scene sketches (e.g., pHash), ambient audio fingerprints, RF beacon environment.
    \item \textbf{Policy (v2):} Admit if \(k\)-of-\(n\) (i) distance bounds validate proximity, (ii) a vision-language model comparison between the disclosed media frame and the witness snapshot yields similarity \(>\tau_{\text{vis}} = 0.70\), and (iii) ambient audio hashes match within tolerance. BLE beacon sets must also overlap.
\end{itemize}

\paragraph{Mobility (Zone Entry/Exit in Urban Transit)}
In mobility applications, the prover is typically a vehicle-mounted unit (e.g., in taxis or delivery vans), and the witnesses are fixed roadside infrastructure. The goal is to produce non-repudiable records of zone crossings or regulated stops, even under GNSS denial or multipath environments.
\begin{itemize}
    \item \textbf{Sensors:} RF environment fingerprints, Doppler shift estimation, IMU bursts, and image hash of nearby infrastructure.
    \item \textbf{Policy (v3):} Admit if \(k\)-of-\(n\) (i) distance bounds validate proximity, (ii) IMU trajectory matches an expected crossing pattern, and (iii) RF fingerprints are consistent with the curbside beacon layout. Optional fallback to rule-based logic applies if visual sensors are unavailable.
\end{itemize}

To make policies concrete, we show a machine-readable example for the Media profile (Figure~\ref{pol:example}). It externalizes quorum and sensing thresholds, is versioned, and is cryptographically bound to the resulting evidence object.

\begin{figure}[h!t]
\centering
\footnotesize
\setlength{\fboxsep}{4pt} 
\begin{boxedminipage}{\dimexpr\columnwidth-6pt\relax}
\hspace{6pt}%
\begin{minipage}{\dimexpr\columnwidth-6pt\relax}
\begin{verbatim}
policy: media_v2
zone_id: Z-17
interval: 2s
quorum:
    k: 3
    n: 4
requirements:
    distance_bound:
        max_distance: 20m
    visual_similarity:
        metric: vlm_embedding
        threshold: 0.70
    audio_hash_match: true
    beacon_overlap:
        min_count: 2
on_fail: reject
\end{verbatim}
\end{minipage}%
\end{boxedminipage}
\vspace{0.6em}
\caption{Illustrative machine-readable admission policy for the Media deployment profile.}
\label{pol:example}
\end{figure}

\subsection{Security and Trust Properties}

The augmented witnessing protocol preserves the spatio-temporal soundness of decentralized PoL systems~\cite{Brito2025SciReports}. Admission still requires time-critical distance bounding with physically proximate witnesses and satisfaction of a $k$-of-$n$ quorum within a fixed block interval $T$, ensuring that accepted claims correspond to bounded events in space and time even under partial collusion, delay, or transient faults.

The augmentation introduces a second, orthogonal protection layer based on verification-time entropy, i.e. unpredictability of witness-local evaluation space conditioned on policy. Beyond proximity, a prover must satisfy contextual coherence checks independently evaluated by multiple witnesses during the same interval. These checks occur locally at capture time under a versioned policy $v$, preventing the prover from predicting which environmental features will be sampled or how they will be assessed. As a result, admissibility cannot be precomputed, and uncertainty is injected directly into the adversarial strategy space.

This entropy derives from \textbf{heterogeneous sensing} modalities, \textbf{policy evolution} across zones and versions, and the \textbf{natural variability} introduced by spatially distributed observers. Even identically equipped witnesses observe distinct environmental slices, making coherent spoofing across a quorum substantially more complex. Fabrication therefore requires not only defeating distance-bounding defenses but also inducing consistent environmental agreement across multiple independent observers within the same temporal window. This significantly increases the cost and complexity of adversarial strategies such as: \begin{itemize} \item \emph{Relay attacks}, in which the prover attempts to satisfy proximity constraints via hidden accomplices; \item \emph{Scene spoofing}, where synthetic or staged environments are constructed to deceive sensors; \item \emph{Sensor spoofing}, where adversaries attempt to replay or predict expected environmental signals; \item \emph{Collusion}, where compromised witnesses are used to fake quorum, mitigated by requiring diversity and independence among participants. \end{itemize}

Accountability is preserved through commitment-based evidence construction. Each admitting witness signs a Merkle root binding its observations, intermediate results, and policy identifier; only such signed commitments participate in quorum formation. Witnesses are assumed to retain the committed descriptors and minimal supporting metadata for a bounded retention period sufficient to enable dispute resolution. The resulting evidence object binds the claim, policy version, and aggregated quorum signature into a verifiable artifact. Incorrect admissions can be challenged post hoc through selective opening of committed elements or reference-policy re-evaluation without disclosing full raw sensor streams or internal model state, maintaining auditability even when local reasoning is approximate or learned.

In sum, the augmented protocol strengthens PoL by combining physical-layer guarantees with strategically unpredictable, sensor-grounded verification while retaining decentralized trust assumptions.

\section{Implementation and Evaluation}
\label{sec:evaluation}
To evaluate resilience against adversarial attempts (e.g., distance fraud) and assess admission policies such as proximity-based and visual verification rules, we implemented a discrete-event simulation using Python and SimPy\footnote{https://simpy.readthedocs.io/en/latest/}. This enables controlled analysis under stochastic environmental conditions and validates protocol behavior and adversarial robustness under an explicitly simplified environment model.

\subsection{Simulation Environment Model}

The simulation implements the witnessing zone model from \Cref{sec:architecture} using discrete events with SimPy. The main modules are:

\begin{itemize}
        \item \textbf{Core Models}: 3D positioning (\texttt{Vector3}), zone configuration (\texttt{ZoneConfig}), and the \texttt{DistanceBoundingProtocol} based on IEEE~802.15.4a\cite{poturalski2011distance} UWB channel models with 32-round exchanges.
    \item \textbf{Environmental Sensing}: Implements \texttt{RfFingerprintSensor} for RSSI-based beacon measurements and \texttt{VisualSensor} for scene object queries.
    \item \textbf{Consensus \& Policy}: It manages \texttt{WitnessNode} voting with a configurable $k$-of-$n$ threshold and evaluates claims against modular policies such as \texttt{SupplyChainPolicy} and \texttt{VisualVerificationPolicy}.
\end{itemize}

The simulation models a single witnessing zone with a radius of $R=20$ meters. The zone is monitored by $N=4$ witness nodes positioned in a square geometry around the zone center. We utilize a quorum threshold of $k=3$, requiring at least three witnesses to independently verify and vote for a location claim before an attestation is issued.
Each simulation experiment runs for a duration of $T=60$ seconds. The prover broadcasts a new location claim every 2 seconds, resulting in a total of $30$ claims per experiment.

The physical layer is modeled using a simplified IEEE 802.15.4a UWB channel model. To account for environmental stochasticity, the path loss ($PL$) at distance $d$ is modeled as:
\begin{equation}
    PL(d) = PL_0 + 10 \gamma \log_{10}\left(\frac{d}{d_0}\right) + X_\sigma
\end{equation}
where $\gamma$ is the path loss exponent ($\gamma=2.0$ for LOS), and $X_\sigma \sim \mathcal{N}(0, \sigma^2)$ represents shadowing effects ($\sigma=3.0$ dB). Ranging errors include a distance-dependent component (1\%) and Gaussian multipath noise.

\subsection{Experimental Scenarios and Admission Policies}

We evaluated the protocol using two policies:

\begin{itemize}
    \item \textbf{Supply Chain Policy:} A proximity-based policy that validates if the prover is within a maximum distance ($2R$) of the witness and possesses an RF fingerprint similarity score above $0.5$. This represents standard logistics tracking.
    \item \textbf{Visual Verification Policy:} Extends the proximity check by requiring witnesses to visually confirm specific semantic queries (e.g., ``Is there a red car?''). This policy fails if the quorum cannot agree on the visual scene description.
\end{itemize}

\begin{figure}[t]
    \centering
    \includegraphics[width=\columnwidth]{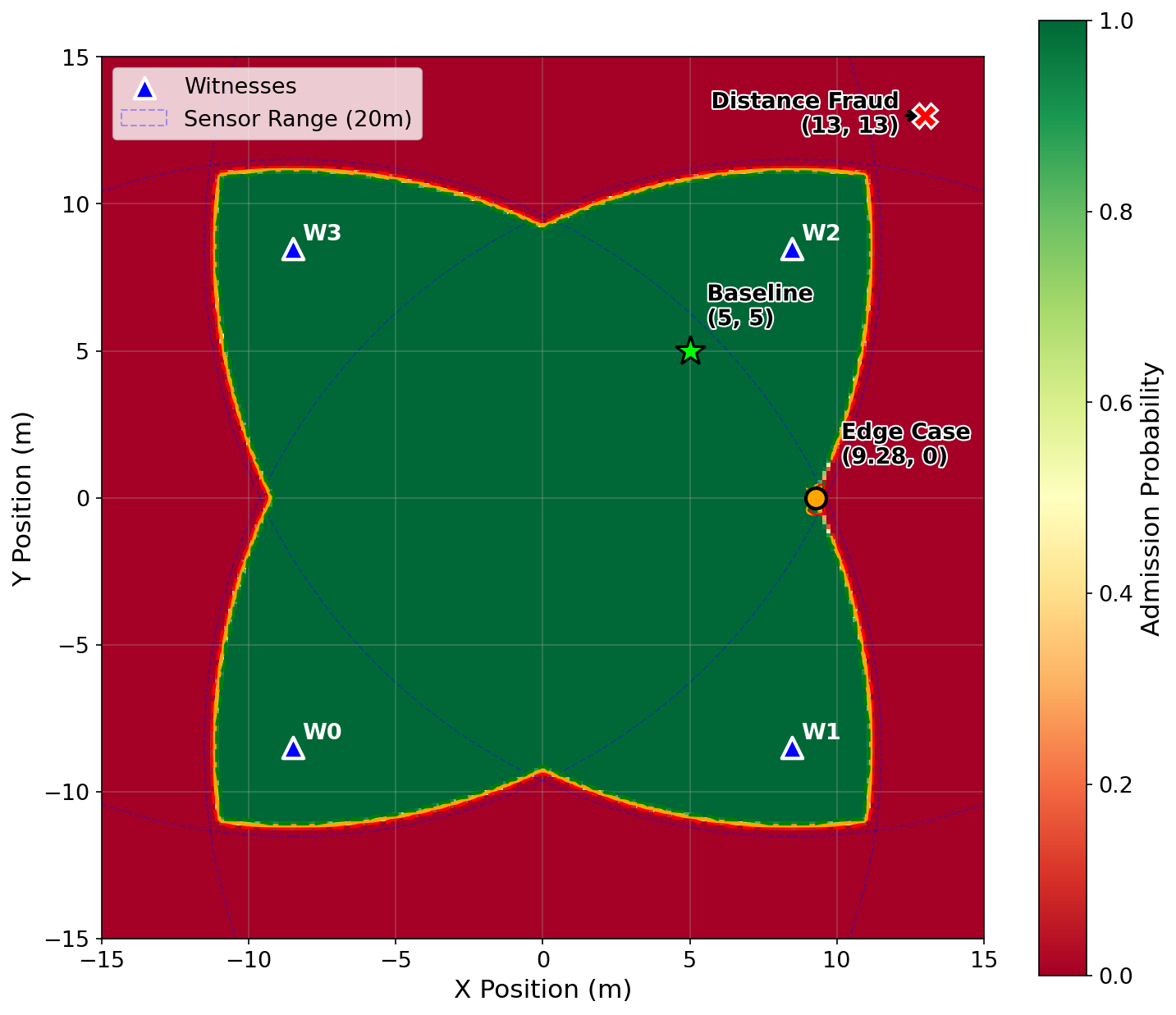}
    \caption{Admission probability heatmap of the simulation environment representing 4W and the theoretical effective witness zone for $d\leq20$ and $k\geq3$.}
    \label{fig:sim-env}
\end{figure}

Using the Supply Chain Policy, we tested four geometric cases, shown in \Cref{fig:sim-env}:

\begin{enumerate}
    \item \textbf{Baseline (4W/6W):} An honest prover is located at $(5, 5)$, well within the zone boundaries. We test with both 4 and 6 witnesses to evaluate the impact of redundancy in the baseline.
    \item \textbf{Distance Fraud:} An external attacker at $(13,13)$ claims to be inside the zone. Since the distance to the farthest witnesses exceeds the acceptance threshold, this scenario tests the protocol's ability to reject spoofed locations.
    \item \textbf{Edge Position:} An honest prover is placed at $(9.28, 0)$, near the effective witnessing boundary. This tests the protocol's sensitivity to edge cases where geometric dilution of precision (GDOP) and shadowing might affect quorum consensus.
\end{enumerate}

\subsection{Performance Analysis}

We conducted a Monte Carlo analysis with 1000 iterations for each scenario to measure the system's reliability in the presence of channel noise. Table~\ref{tab:scenario_comparison} summarizes the performance across the four geometric scenarios\footnote{N/A as no positive instances exist (TP+FP=0, TP+FN=0).}.

The \textbf{Baseline (4W)} scenario demonstrates a 100\% success rate for honest users within the zone. Increasing the number of witnesses to 6 (\textbf{Baseline (6W)}) maintained this performance, confirming that the witness configuration provides robust coverage.

\subsection{Security Analysis}

The \textbf{Distance Fraud} scenario confirms the protocol's security. Attackers positioned outside the zone attempting to spoof closer positions were rejected in 100\% of cases.

The \textbf{Edge Position} scenario highlights the conservative nature of the protocol. Users located exactly at the zone boundary experience a significant drop in admission rate (35.9\%), indicating that the system prefers to reject ambiguous edge cases rather than admit potential attackers.

In addition to RF constraints, the \textit{VisualVerificationPolicy} achieved a high admission rate when the required "Red Car" object was present (97\%) and dropped to 0\% when the object was removed, showing that the witnessing consensus correctly enforced environmental prerequisites.

This evaluation is intentionally limited to simulation and should be read as a feasibility check rather than operational proof. It does not measure hardware timing, field latency, ledger finality, or deployment-scale throughput. Relative to baseline decentralized PoL costs already analyzed in prior work~\cite{Brito2025SciReports,brito2025taxonomy}, the augmentation adds witness-local sensing, one commitment root per admitting witness, and policy-dependent computation whose cost scales mainly with the selected modalities and quorum size rather than with new global protocol rounds.

\begin{table}[h]
    \centering
    \caption{Comparative Analysis of Simulation Scenarios (1000 iterations)}
    \label{tab:scenario_comparison}
    \begin{tabular}{lcccc}
        \toprule
        \textbf{Scenario} & \textbf{Success Rate} & \textbf{Precision} & \textbf{Recall} & \textbf{Admitted} \\
             \midrule
        Baseline (4W)      & 1.000 $\pm$ 0.00 & 1.00 & 1.000 & 30.0/30 \\
        Baseline (6W)      & 1.000 $\pm$ 0.00 & 1.00 & 1.000 & 30.0/30 \\
        Distance Fraud     & 0.000 $\pm$ 0.00 & N/A & N/A   & 0.0/30  \\
        Edge Position      & 0.359 $\pm$ 0.09 & 1.00 & 0.359 & 10.8/30 \\
        \midrule
        Visual (Valid)     & 0.973 $\pm$ 0.08 & 1.00 & 0.973 & 29.2/30 \\
        Visual (Invalid)   & 0.000 $\pm$ 0.00 & N/A & N/A & 0.0/30 \\
        \bottomrule
    \end{tabular}

\end{table}

\section{Discussion}
\label{sec:discussion}
The augmented witnessing-zone architecture should be understood as a generalizable design pattern rather than a domain-specific solution. The core construct applies to settings where disputes concern not only \emph{where} and \emph{when}, but also \emph{what} occurred. DT engineering is one natural beneficiary, since the evidence object can serve as a verifiable admission layer for externally sourced observations before they influence twin state, automation, or audit records, but the same interface is relevant to other multi-stakeholder cyber-physical systems. Open questions remain around policy lifecycle management, cross-zone interoperability, evidence portability for long-lived DTs, and adversarial behavior beyond single-zone distance fraud. The current evaluation provides initial validation rather than operational proof, and privacy remains a deployment trade-off~\cite{bogdanov2025zero}: although the evidence object exposes commitments rather than raw sensor streams, witnesses still observe local context and selective opening may reveal sensitive environmental features. Hardware-in-the-loop experiments, field pilots, privacy-preserving policy verification, and scalability analysis across federated zone topologies are therefore natural next steps.

\section{Conclusion}
\label{sec:conclusion}
This paper addressed an important challenge for cyber-physical systems operating in distributed, multi-stakeholder settings with externally sourced observations: ensuring that such data is grounded in verifiable physical reality at capture time. Building on Proof-of-Location as a baseline for establishing where and when events occur, we introduced an augmented witnessing-zone architecture in which multiple independent observers collectively validate physical events and produce auditable evidence objects for downstream verification. By extending spatio-temporal attestation with contextual witness agreement, the approach strengthens resistance to scene spoofing and other forms of fabricated or staged evidence at capture time. The result is an architectural building block for trustworthy data admission in settings where externally sourced observations influence system state, analytics, or audit records. The evaluation shows that this can be achieved under the studied simulation assumptions without undermining operational predictability.

\bibliographystyle{IEEEtran}
\bibliography{ref}

\end{document}